\documentclass[aps,prc,twocolumn,10pt,superscriptaddress,preprintnumbers,showpacs,nofootinbib,a4paper]{revtex4-2}

\usepackage{color,amssymb,graphicx,isotope}
\usepackage[linkcolor=blue,citecolor=blue,urlcolor=blue,colorlinks=true]{hyperref}

\usepackage[range-units=single,separate-uncertainty=true,separate-uncertainty-units=single]{siunitx}
\usepackage[version=4]{mhchem}
\usepackage[capitalise]{cleveref}
\graphicspath{{../figures/}{figures/}}

\usepackage[switch]{lineno}
\modulolinenumbers[5]

\newcommand{\tu}{\affiliation{Institut f\"ur Kern- und Teilchenphysik (IKTP), Technische Universit\"at Dresden, Dresden, Germany}}
\newcommand{\hzdra}{\affiliation{Institut für Ionenstrahlphysik und Materialforschung, Helmholtz-Zentrum Dresden-Rossendorf (HZDR), Dresden, Germany}}
\newcommand{\hzdrb}{\affiliation{Institut für Strahlenphysik, Helmholtz-Zentrum Dresden-Rossendorf (HZDR), Dresden, Germany}}
\newcommand{\france}{\affiliation{Present address: Department of Physics, University of Toronto, Toronto, Canada}}

\begin{document}

\title{Evidence for neutron-induced $\boldsymbol{\gamma}$-ray emissions in the vicinity of the $\boldsymbol{Q}$ value of \ce{^76Ge} $\boldsymbol{0\nu\beta\beta}$ decay}
	
\date{\today}
\author{Marie Pichotta}
\email[Contact author: ]{marie.pichotta@tu-dresden.de}
\tu
\author{Toralf Döring}\tu \hzdra
\author{Hans~F.~R. Hoffmann}\tu 
\author{Konrad Schmidt}\hzdrb
\author{Ronald Schwengner}\hzdrb
\author{Steffen Turkat}\tu 
\author{Birgit Zatschler}\tu \france 
\author{Kai Zuber}\tu

\begin{abstract}
	Neutrinoless double-beta decay of nuclei represents one of the most promising methods for uncovering physics beyond the Standard Model. In this context, \ce{^{76}Ge} stands out as a particularly attractive candidate, as it can serve as an intrinsic component in semiconductor detectors. If the neutrinoless process occurs in \ce{^{76}Ge}, its signature would appear as a distinct peak at the $Q$ value of \qty{2039}{\keV}. A neutron activation measurement was performed on a germanium sample isotopically enriched in \ce{^76Ge} at the DT neutron generator of TU Dresden. The measurement confirmed the presence of $\gamma$ rays with energies of \qty{2033.1+-0.5}{\keV}, \qty{2035.5+-0.4}{\keV}, and \qty{2040.22+-0.26}{\keV} originating from the decays of \ce{^74Ga} and \ce{^76Ga}. These $\gamma$ rays lie in close proximity to the expected neutrinoless double-beta decay signal of \ce{^76Ge}.
\end{abstract}
	
\maketitle

\section{Introduction}
Although the Standard Model provides a robust framework for understanding fundamental particles and their interactions, several neutrino-related issues remain unresolved, which are critical to advancements in neutrino, nuclear, particle, and astroparticle physics \cite{Ejiri2019PRP}.
One open question is the mechanism by which neutrinos acquire mass. Neutrino oscillations have indirectly confirmed that neutrinos are massive \cite{Fukuda1998PRL,Ahmad2002PRL,Ahmed2002PRL}, but the underlying process responsible for generating their mass remains unresolved. Moreover, the existence of sterile neutrinos and the potential for ultra-heavy neutrinos beyond the \mbox{Standard} Model are still subjects of active investigation. Finally, the determination of the absolute neutrino mass scale continues to be an experimental challenge, with various efforts underway to address this fundamental uncertainty.

In particular, the observation of neutrinoless double-beta ($0\nu\beta\beta$) decay, a process that violates lepton number ($\Delta L = 2$), could provide crucial insights. Detecting this decay would directly verify that neutrinos are of the Majorana type, and hence their own antiparticles, providing compelling evidence for physics beyond the \mbox{Standard Model \cite{Dolinski2019ARNPS}}. To date, experiments probing \(0\nu\beta\beta\) decay (including KamLAND-Zen \cite{Abe2023PRL}, \mbox{CUORE \cite{Adams2022NAT}}, GERDA \cite{Agostini2020PRL}, MAJORANA \cite{Arnquist2023PRL}, EXO-200 \cite{Anton2019PRL}, and CUPID-Mo \cite{Augier2022EPJC}) have established lower half-life limits exceeding \qty{e26}{years}. Consequently, the next generation of $0\nu\beta\beta$ decay experiments will require considerably larger quantities of isotopically enriched material (on the order of tons) while simultaneously reducing background rates. Several upcoming projects aim to achieve multi-ton scale detectors, such as LEGEND \cite{Abgrall2021arxiv}, nEXO \cite{Adhikari2022JPG}, SNO+ \cite{Albanese2021JINST}, JUNO \cite{An2016JPG}, CDEX-1B \cite{Zhang2023arxiv}, THEIA \cite{Askins2020EPJC}, DarkSide-20k \cite{Aalseth2018EPJP}, and DARWIN \cite{Agostini2020EPJC}.

In total, \qty{35}{nuclides} could potentially undergo \mbox{double} $\beta^-$ decay. However, \ce{^76Ge} is particularly well-suited for experiments that employ the source-as-detector approach. This is primarily because high-purity germanium (HPGe) detectors can be readily enriched in \ce{^76Ge} and offer excellent energy resolution. These advantages have been utilized by the LEGEND experiment \cite{Abgrall2021arxiv}, which builds upon the infrastructure of the previous \ce{^76Ge} $0\nu\beta\beta$ decay experiments GERDA \cite{Agostini2020PRL} and MAJORANA \cite{Arnquist2023PRL}, and targets a total germanium detector mass of \qty{1000}{\kg}. The signature of \ce{^76Ge} $0\nu\beta\beta$ decay is characterized by a peak at its $Q$ value of \qty{2039.06+-0.01}{\keV} \cite{Wang2021CPC}. A thorough understanding of the background in the vicinity of \qty{2039}{\keV} is essential for excluding false event triggers. While \mbox{LEGEND} uses a fit window from \qtyrange{1985}{2095}{\keV}, centered on the $Q$ value, to determine the background indices \cite{Abgrall2021arxiv}, the present study focuses on the spectral region from \qtyrange{1980}{2065}{\keV}.

\section{State of the art}
Although LEGEND employs advanced active veto systems and passive shielding, cosmic muons can still traverse the detector system undetected, thereby generating neutrons across a broad energy spectrum capable of interacting with the germanium \cite{Mei2006PRD}. As the detectors in LEGEND are predominantly composed of \ce{^74Ge} (averaging 12.4\% in GERDA and 8.65\% in \mbox{MAJORANA \cite{Ackerman2013EPJC}}) and \ce{^76Ge} (averaging 87.4\% in GERDA and 91.4\% in \mbox{MAJORANA} \cite{Ackerman2013EPJC}), the principal source of in-situ cosmogenic background stems mainly from muon-induced neutrons interacting with these isotopes \cite{Abgrall2021arxiv}. Indeed, prior studies on the neutron activation of isotopically enriched germanium have revealed several $\gamma$ rays with energies near the region of interest for \ce{^76Ge} 0$\nu\beta\beta$, potentially mimicking the signal of this decay.

\subsection{Neutron activation populating \ce{^74Ge}}
For instance, Camp, Fielder, and Foster reported a $\gamma$-ray peak at \qty{2036.20+-0.37}{\keV} with a relative intensity of \mbox{0.19 $\pm$ 0.04\%} \cite{Camp1971NPA1}. This transition was observed during the $\beta^-$ decay of \ce{^74Ga}, which populates a \qty{4201.39}{\keV} level in \ce{^74Ge} that subsequently de-excites via the emission of the $\gamma$ ray. In their experiment, isotopically enriched \ce{GeO2} samples (comprising 94.5\% \ce{^74Ge}), each weighing \qty{60}{\mg}, were irradiated with a neutron flux of $10^{11}$ to $10^{12}$\,$n$\,cm$^{-2}$\,s$^{-1}$ at \qty{14}{\MeV}. After each irradiation, the gallium produced via the \ce{^74Ge($n$,$p$)^74Ga} reaction was chemically separated and quantified using a \ce{Ge(Li)} detector with a relative efficiency of 3.8\%.

Taylor \textit{et al.} \cite{Taylor1975CJP} verified this transition by irradiating \qty{1}{g} of isotopically enriched \ce{GeO2} (comprising 94.48\% \ce{^74Ge}) with \qty{14.7}{\MeV} neutrons. Using a \ce{Ge(Li)} detector with 9.5\% relative efficiency, they measured a $\gamma$-ray energy of \qty{2036.2+-0.9}{\keV} and a relative intensity of \mbox{0.13 $\pm$ 0.08\%}, in agreement with the findings of Camp and Foster \cite{Camp1971NPA1}. Combined, these results yield a $\gamma$-ray energy of \qty{2036.2+-0.4}{keV} and an emission probability of 0.17 $\pm$ 0.04\% \cite{Singh2006NDS}.

\subsection{Neutron activation populating \ce{^76Ge}}
In another experiment, Camp and Foster activated two isotopically enriched \ce{GeO2} samples (comprising 95.16\% \ce{^76Ge}), each weighing \qty{105}{\mg}, via the \ce{^76Ge($n$,$p$)^76Ga} reaction. The neutron beam and counting detector were identical to those employed in the \ce{^74Ga} study. In this configuration, they observed a $\gamma$-ray peak at \mbox{\qty{2040.70+-0.25}{\keV}} with a relative intensity of \mbox{0.50 $\pm$ 0.08\%} \cite{Camp1971NPA2}, corresponding to an emission probability of 0.29 $\pm$ 0.05\% \cite{Singh2024NDS}. This $\gamma$-ray transition occurred during the de-excitation of a \qty{3951.89}{\keV} level in \ce{^76Ge}, which was populated by the decay of \ce{^76Ga}.

However, Rouki \textit{et al.} did not observe this transition. They irradiated two isotopically enriched germanium samples with masses of \qty{14.56}{\g} (comprising 12.3\% \ce{^74Ge} and 87\% \ce{^76Ge}) and \qty{17.43}{\g} (containing 10.350\% \ce{^74Ge} and 87.44\% \ce{^76Ge}) using a continuous-energy neutron beam with energies up to \qty{20}{\MeV}, produced by the GELINA white neutron source \cite{Rouki2013PRC}. In this experiment, the objective was to populate the excited states of \ce{^76Ge} via \ce{($n$,$n${'}\gamma)} reactions, with data acquired using the GAINS spectrometer equipped with eight HPGe detectors.

Domula \textit{et al.} extended the search for the \qty{2040.70}{\keV} $\gamma$-ray by activating the \qty{17.43}{\g} \ce{Ge} crystal previously utilized in the Rouki experiment. This activation employed \qty{14}{\MeV} neutrons from the DT neutron generator at TU Dresden \cite{Domula2014NDS}, with the ensuing measurements performed using two HPGe detectors with relative efficiencies of 25\% and 30\% \cite{DomulaPhD}. Although the resulting spectrum exhibited a peak-like structure in the region of interest, an unbiased fit could not be performed owing to the limited energy resolution.

A separate investigation of the $\gamma$-ray transitions emanating from the \qty{3951.89}{\keV} excited state of \ce{^76Ge}, induced via inelastic neutron scattering \ce{($n$,$n${'}\gamma)} by the Kentucky group \cite{Crider2015PRC}, failed to detect the \qty{2040.70}{\keV} $\gamma$-ray. In the experiment, neutrons with energies up to \qty{4.9}{\MeV} were produced via the \ce{^3H($p$,$n$)^3He} reaction using protons from a 7-MV Van de Graaff accelerator. Two different targets were irradiated: an isotopically enriched \ce{Ge} powder (84.12\% \ce{^76Ge}, \qty{11.13}{\g}) and an isotopically enriched \ce{GeO2} powder (85\% \ce{^76Ge}, \qty{41.84}{\g}). The measurements were performed with a single HPGe detector, which was surrounded by a bismuth germanate (BGO) detector for Compton suppression of scattered $\gamma$ rays. Nonetheless, they reported evidence for a new excited state in \ce{^76Ge} at \qty{3147}{\keV} that de-excites by emitting a $\gamma$ ray of \qty{2037.5+-0.3}{\keV}. Moreover, they attributed an additional contribution to the \qty{2037.5}{\keV} peak to a possible transition from a newly identified \qty{3577}{\keV} level to a lower-lying \qty{1539}{\keV} state. In a subsequent publication, this group refined the energy of the \qty{3147.28+-0.13}{\keV} level, yielding a revised $\gamma$-ray energy of \mbox{\qty{2038.89+-0.15}{\keV} \cite{Mukhopadhyay2017PRC}}.

Tornow \textit{et al.} investigated $\gamma$-ray transitions in \ce{^76Ge} induced by \ce{($n$,$p$)} reactions by activating \qty{3.2}{\g} of isotopically enriched germanium (comprising 14\% \ce{^74Ge} and 86\% \ce{^76Ge}) with neutrons up to \qty{20}{\MeV} from the TUNL neutron source, applying irradiation of \qty{120}{\s} per cycle \cite{Tornow2016PRC}. A HPGe detector with 60\% relative efficiency, enclosed within lead shielding, was utilized for the measurements. The summed spectrum from \num{110} irradiation cycles revealed a peak-like structure near \qty{2039.4}{\keV}. However, an unbiased fit could not be performed due to the high spectral background and insufficient statistics.

\subsection{Further Approaches}
An alternative method for populating the excited states of \ce{^76Ge} was pursued by Schwengner \textit{et al.} They employed bremsstrahlung photons generated by electron beams of \qtylist{7.8;12.3}{\MeV} from the ELBE accelerator to induce \ce{(\gamma,\gamma{'})} reactions \cite{Schwengner2022}. The target comprised \qty{1.8760}{\g} of isotopically enriched germanium (containing 93.4\% \ce{^76Ge}), and the detection system utilized four HPGe detectors, each with 100\% relative efficiency, arranged at different angles. Nevertheless, the resulting spectrum did not exhibit a clear indication of the \qty{2040.70}{\keV} $\gamma$-ray transition. This absence might be ascribed either to alternative spin configurations differing from 1 or to insufficient feeding of the \qty{3951.89}{\keV} level from higher-lying states.

It is noteworthy that previous experiments have also reported both prompt and delayed $\gamma$-ray transitions in \ce{^77Ge} induced by \ce{($n$,\gamma)} reactions \cite{Lent1974PhD,Meierhofer2012EPJA}. However, given the limited sensitivity of the present experiment to this channel, these $\gamma$ rays will not be considered further.

The aim of this work is to elucidate the potential $\gamma$-ray background present near \qty{2039}{\keV} in isotopically enriched germanium. To this end, $\gamma$-ray spectra from an isotopically enriched \ce{Ge} sample and from a sample with natural composition activated by fast neutrons have been analyzed in detail.

\section{Neutron Irradiation Procedure}
Neutron irradiation was conducted at the DT neutron generator of TU Dresden, located at Helmholtz-Zentrum Dresden-Rossendorf (HZDR) \cite{Klix2018EPJWC}. In this setup, a primary deuterium beam from a duoplasmatron ion source is accelerated by a Cockcroft-Walton type accelerator operating at a current of several \unit{\mA}. The beam is focused onto a water-cooled target consisting of a copper disk overlaid with a titanium layer implanted with tritium (approximately \qty{0.7}{\tera\becquerel}). As a result of the fusion reaction \ce{^3H(^2H,$n$)^4He}, neutrons with a discrete energy of \qty{14.1}{\MeV} and an intensity of up to $10^{12}$\,$n$\,s$^{-1}$ are produced.

The neutron beam irradiated a conically shaped, monocrystalline \ce{Ge} sample with a mass of \qty{17.42}{\g}, a height of \qty{25.6}{\mm}, and diameters of \qty{7.0}{\mm} at the apex and \qty{22.5}{\mm} at the base. This sample, one of the cusps extracted from the cylindrical ingots used to fabricate the detectors for the GERDA experiment, maintains the same isotopic composition: 0.001\% \ce{^70Ge}, 0.027\% \ce{^72Ge}, 0.110\% \ce{^73Ge}, 12.350\% \ce{^74Ge}, and 87.512\% \ce{^76Ge} \cite{Ackerman2013EPJC}. For comparison, a \qty{23.40}{\g} \ce{Ge} crystal of natural abundance (comprising 20.5\% \ce{^70Ge}, 27.4\% \ce{^72Ge}, 7.8\% \ce{^73Ge}, 36.5\% \ce{^74Ge}, and 7.44\% \ce{^76Ge}) was used (see \cref{fig:figure1}).

The samples were irradiated in multiple cycles, with an activation time of \qty{140}{\s} chosen to optimize both the peak-to-Compton ratio and the counting statistics in the $\gamma$-ray peaks near \qty{2039}{\keV} in the spectrum of the isotopically enriched \ce{Ge} sample.

During irradiation, the sample was positioned in close proximity to the neutron-production target, within a foam-lined transportation shuttle made of borated polyethylene, and integrated into a pneumatic tube system. Polyethylene was chosen for its stability, which minimizes the required material and reduces parasitic activation. Furthermore, the incorporation of boron enhances neutron capture via the \ce{^10B($n$,\alpha)^6Li} reaction, mitigating the impact of thermal neutrons generated by the hydrogen in polyethylene. However, due to the low amount of material, this effect is rather negligible.

After irradiation, the sample was transferred to the counting station, arriving at the tube system's end piece constructed from acrylic glass and aluminum. Photoelectric sensors provided position information and generated start and stop signals for each irradiation and transfer cycle. The transfer itself took approximately \qty{11}{\s}. The prompt increase in the count rate observed in the HPGe detectors then marked the onset of the decay measurement at its maximum rate.

\section{Detection Setup}
\begin{figure}
	\includegraphics[width=\linewidth]{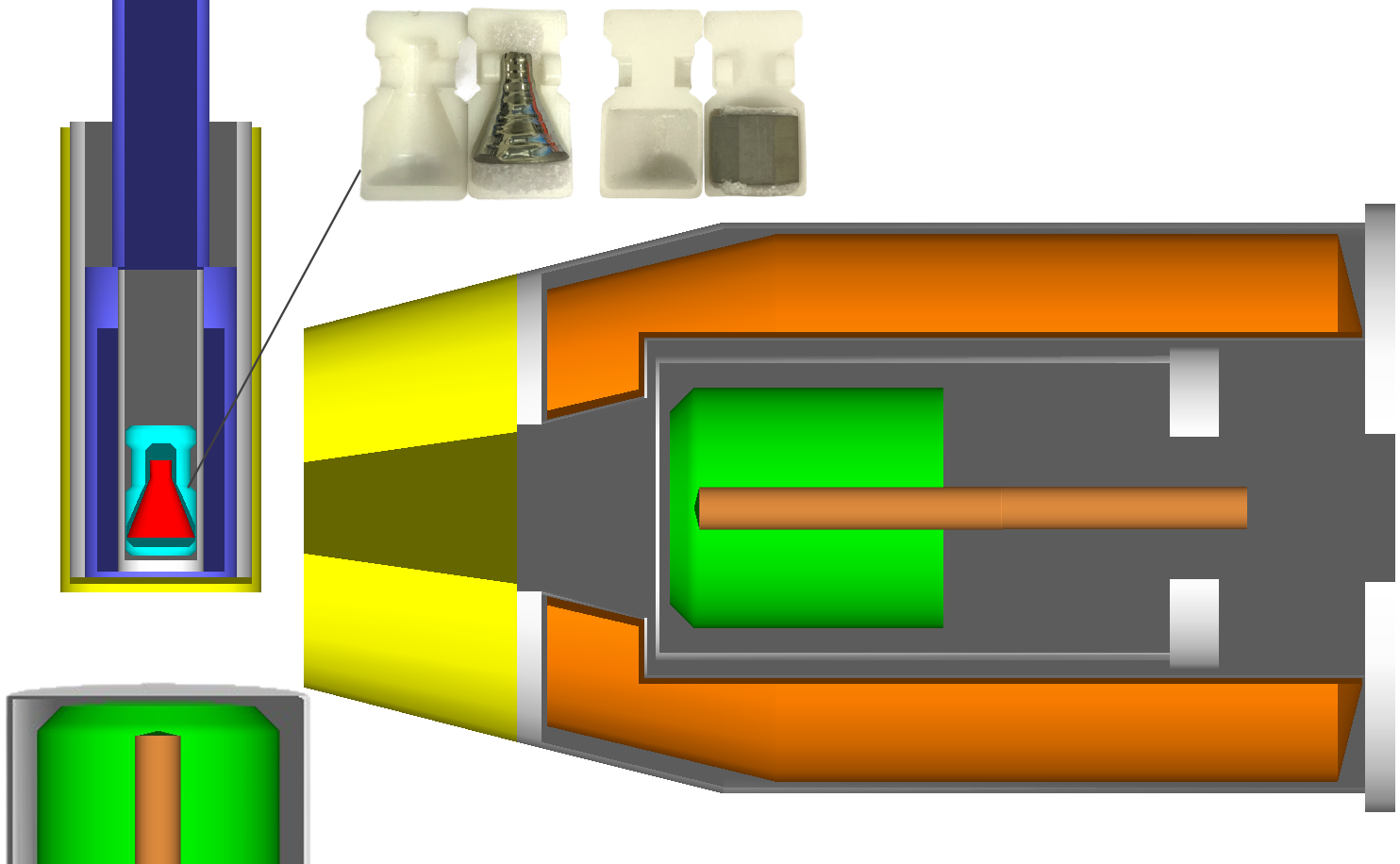}%
	\caption{\label{fig:figure1} Schematic illustration of the counting setup. The isotopically enriched \ce{Ge} sample (red) is housed within its transportation shuttle, made of borated polyethylene (cyan). This shuttle is positioned inside the end piece of the pneumatic tube, which is constructed from acrylic glass (blue) and aluminum (gray). The entire assembly is further enclosed by a \qty{5}{\mm} thick aluminum layer (gray) and a \qty{3}{\mm} thick lead layer (yellow). The $\gamma$ rays are detected by HPGe detector D1 (active volume highlighted in green), which is surrounded by a BGO detector (orange) for Compton suppression. A lead collimator (yellow) shields the BGO detector to block radiation emanating directly from the sample. Additionally, a second HPGe detector D2 is aligned beneath the sample. Additionally, photographs of the germanium samples, one isotopically enriched in \ce{^76Ge} (left) and one with natural composition (right), are shown within their respective transportation shuttles.}
\end{figure}

A \qty{70}{\mm} thick lead collimator was positioned in front of a \qty{30}{\mm} thick BGO detector, which served both to suppress Compton scattering and to reduce single and double escape peaks in the spectra of an n-type HPGe detector (D1) with a relative efficiency of 110\% (see \cref{fig:figure1}). The central axes of the collimator, the BGO detector, and D1 were aligned with the geometrical center of the activated sample, which was housed within the transportation shuttle. The distance between the detector end cap and the sample center was \qty{16.5}{\cm}. In the initial phase of the experiment, a second n-type HPGe detector (D2) with a relative efficiency of 100\% was also implemented, with its crystal positioned vertically below the \ce{Ge} sample at a distance of \qty{6.5}{\cm}. However, due to the shorter distance to the source and the absence of Compton suppression, the spectra obtained from D2 exhibit poorer energy resolution and higher background in the region of interest.

The $\gamma$-ray signals from the HPGe detectors were recorded in list mode using a CAEN DT5720 digitizer. This method ensured that each event was individually logged along with attendant parameters, such as energy and timestamps.

Additionally, the counting setup was optimized through Geant4 simulations \cite{Agostinelli2003NIMA, Allison2006ITNS, Allison2006NIMA} within the MaGe framework \cite{Boswell2011ITNS}. The customized physics lists, which incorporate radioactive decay processes and utilize the unpolarized \mbox{Livermore} low-energy model, are particularly well suited for low-background applications. In the simulation, the sample was modeled as a germanium volume with an isotopic composition identical to that of the isotopically enriched \ce{Ge} sample. Decays of both \ce{^74Ga} and \ce{^76Ga} were initiated within this volume, with the resulting energy depositions in the sensitive detector regions being recorded on an event-by-event basis.

Simulations of various setup configurations confirmed that electrons produced in the $\beta^-$ decays are predominantly attenuated via ionization rather than by bremsstrahlung when low-Z materials, such as borated polyethylene (used for the shuttle), acrylic glass, and aluminum (used for the tube), are employed. For this reason, a \qty{5}{\mm} thick aluminum layer was additionally incorporated around the tube to further suppress the generation of bremsstrahlung. Furthermore, random coincidences caused by the high count rate may also contribute to an increased background in the region of interest. In particular, \ce{^75Ge} and \ce{^{75\textit{m}}Ge}, produced via the \ce{^74Ge($n$,\gamma)} and \ce{^76Ge($n$,{2}$n$)} reactions, respectively, give rise to the most intense $\gamma$-ray peaks in the isotopically enriched \ce{Ge} sample spectrum at \qty{264.6}{\keV} and \mbox{\qty{139.68}{\keV} \cite{Negret2013NDS}}. Consequently, a secondary shield consisting of \qty{3}{\mm} thick lead was employed. This shielding effectively attenuates low-energy photons while leaving $\gamma$ rays above \qty{2}{\MeV} nearly unaffected. In this way, any low-energy bremsstrahlung was further reduced and the overall count rate was lowered, thereby minimizing the likelihood of dead time in the readout electronics (see \cref{fig:figure1}).

\section{Data Processing workflow}
Since the spectra were acquired in list mode, the data were analyzed using distinct timing windows. This strategy enabled the identification of the origins of $\gamma$-ray peaks by exploiting the different half-lives of their parent nuclides, while also enhancing the peak-to-Compton ratio near \qty{2039}{\keV}. The peak-to-Compton ratio is mainly degraded by additional decays (e.g., \ce{^74Ga} with \mbox{$T_{1/2}=\qty{487+-7}{\s}$} \cite{Singh2006NDS}) that occur simultaneously with the decay of interest (e.g., \ce{^76Ga} with \mbox{$T_{1/2}=\qty{30.5+-0.6}{\s}$} \cite{Singh2024NDS}), thereby contributing to the Compton-induced background.

\begin{figure*}
	\includegraphics[width=\textwidth, trim=7mm 3mm 3mm 5.5mm]{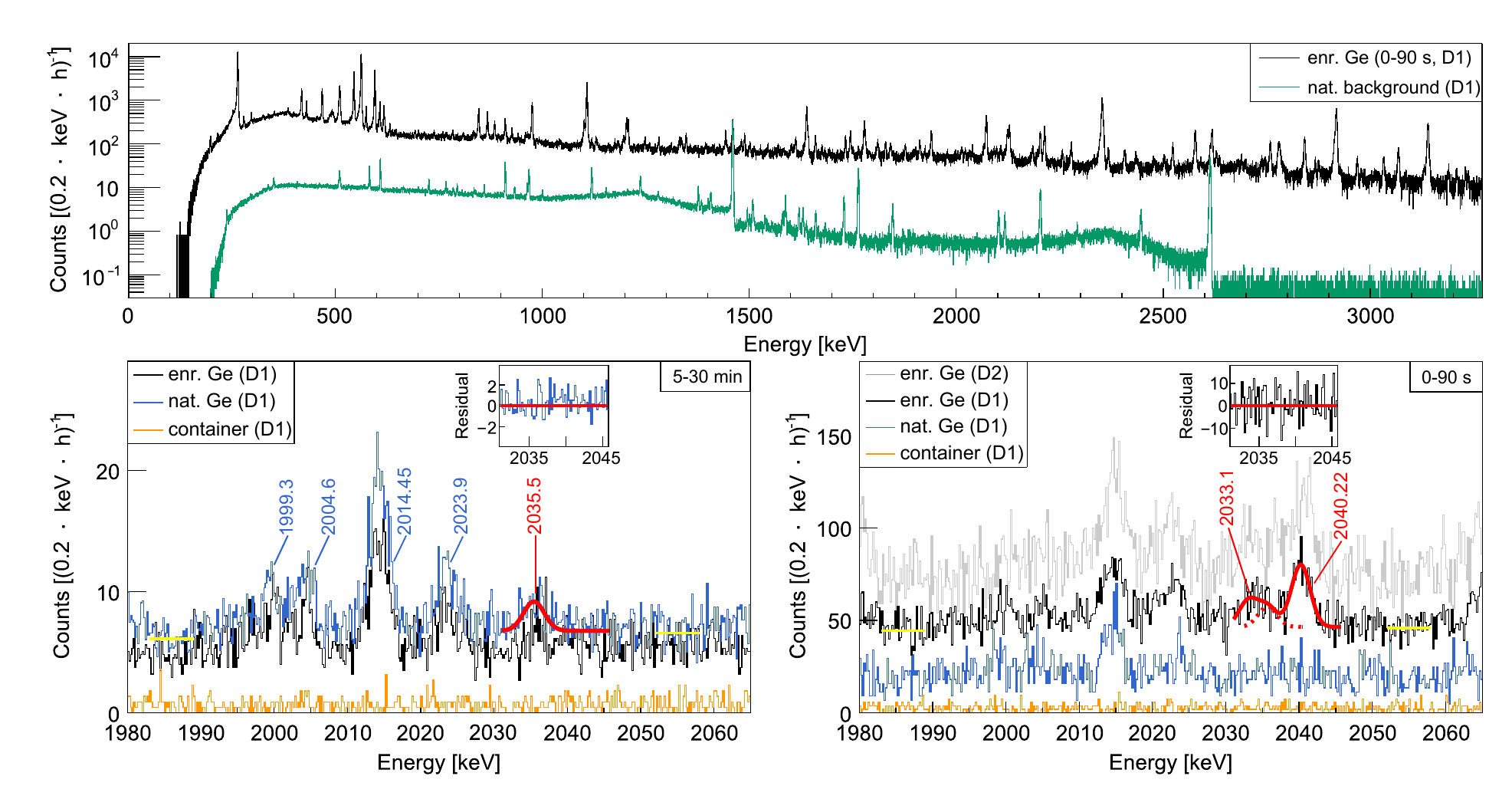}%
	\caption{\label{fig:figure2} \textbf{Top:} The count rate of the natural laboratory background (green), measured continuously over \qty{43.7}{\hour}, is compared with that of the isotopically enriched \ce{Ge} sample (black, \num{50} runs) recorded in a timing window of \qtyrange{0}{90}{\s}. \textbf{Bottom left:} In the energy range \qtyrange{1980}{2070}{\keV}, the count rate of the isotopically enriched \ce{Ge} sample (black, \num{9} runs) is compared to the summed spectrum of the natural \ce{Ge} sample (blue, \num{4} runs) and that of the activated empty transportation shuttle (orange, \num{5} runs) near the region of interest for a timing window of \qtyrange{5}{30}{\minute}. The $\gamma$-ray peak at \qty{2035.5}{\keV} in the spectrum of the natural \ce{Ge} sample is fitted with a single Gaussian (red) based on constant background fits (yellow), and visible $\gamma$-ray peaks from \ce{^74Ga} are labeled in blue. \textbf{Bottom right:} The count rate of the isotopically enriched \ce{Ge} sample (black, \num{50} runs) is compared with that of the natural \ce{Ge} sample (blue, \num{4} runs), the empty transportation shuttle (orange, \num{13} runs), and the spectrum recorded by HPGe detector D2 (gray, \num{40} runs) for a timing window of \qtyrange{0}{90}{\s}. To enhance visibility, the gray spectrum is shifted by \num{-50} counts per (0.2$\cdot$\,keV$\cdot$\,h). In this panel, the $\gamma$-ray peaks at \qtylist[list-final-separator = {, and }]{2033.1;2035.5;2040.22}{\keV} in the spectrum of the isotopically enriched \ce{Ge} sample are approximated using a triple Gaussian fit (red) on top of constant background fits (yellow). For both Gaussian approximations, the residuals are plotted to represent the deviation of the data from the fitting functions.}
\end{figure*}

For the black spectrum in \cref{fig:figure2} (top) and for all spectra shown in \cref{fig:figure2} (bottom right), counting began immediately upon the samples' arrival at the counting station and continued for \qty{90}{\s} (timing window \qtyrange{0}{90}{\s}). This measurement interval was chosen to optimize the peak-to-Compton ratio near \qty{2039}{\keV} in the spectrum of the isotopically enriched \ce{Ge} sample by primarily excluding events induced by \ce{^74Ga}. Conversely, in \cref{fig:figure2} (bottom left), the samples were allowed to decay for \qty{5}{\minute} before a \qty{25}{\minute} measurement (timing window \qtyrange{5}{30}{\minute}) was performed. This delay ensured that most of the short-lived nuclides have decayed, thereby significantly reducing their contribution to the Compton-induced background.

The $\gamma$-ray spectra of both the isotopically enriched and natural \ce{Ge} samples were energy-calibrated in situ. For the \qtyrange{0}{90}{\s} timing window, calibration relied on the three high-intensity $\gamma$-ray transitions from \ce{^76Ga} at \mbox{\qty{545.51+-0.03}{\keV}}, \qty{1108.41+-0.08}{\keV}, and \mbox{\qty{2919.85+-0.10}{\keV}} \cite{Singh2024NDS}. In contrast, for the \qtyrange{5}{30}{\minute} timing window the calibration was based on the \ce{^74Ga} $\gamma$-rays observed at \qty{1489.37+-0.07}{\keV}, \qty{1940.63+-0.07}{\keV}, and \qty{2279.05+-0.09}{\keV} \cite{Singh2006NDS}. Each individual spectrum was calibrated separately to ensure an accurate summation, and the final results incorporate the uncertainties associated with this calibration process.

Furthermore, the in-beam spectra are significantly influenced by fluctuations in the neutron flux from the DT neutron generator, which directly affect the activities of the radionuclides. To facilitate a meaningful comparison of the spectra presented in \cref{fig:figure2}, the spectra of the natural \ce{Ge} sample and the empty transportation shuttle are normalized to their respective isotopically enriched \ce{Ge} sample spectrum measured by detector D1. This normalization is achieved by comparing the intensity of the \qty{1778.987}{\keV} $\gamma$-ray peak from \ce{^28Al} \cite{Singh2013NDS}, which originates from activation of the transportation shuttle and serves as a proxy for the neutron flux during irradiation.

\section{Analysis of the region of interest}
The resulting count rate per \qty{0.2}{\keV}-wide channel, obtained from \num{50} irradiation cycles (timing window \qtyrange{0}{90}{\s}) of the isotopically enriched \ce{Ge} sample measured by the HPGe detector D1, is shown as the black spectrum in \cref{fig:figure2} (top) and is compared with the naturally occurring laboratory background (green). In this short-duration spectrum, the dominant $\gamma$-ray features arise from \ce{^74Ga} and \ce{^76Ga}. Additionally, full-energy peaks corresponding to \ce{^71Zn}, \ce{^73Zn}, \ce{^75Ga}, \ce{^75Ge}, \ce{^{75{m}}Ge}, and \ce{^{77{m}}Ge} are visible, originating from \ce{($n$,{2}$n$)}, \ce{(n,\gamma)}, \ce{($n$,\alpha)}, \ce{($n$,$np$)}, and \ce{($n$,$d$)} reactions. Moreover, $\gamma$-ray signals from \ce{^11Be} and \ce{^28Al} appear in the spectrum due to parasitic activation of the transportation shuttle and the cushioning material.\\

For the investigation of the $\gamma$-ray peaks induced by \ce{^74Ga} near \qty{2039}{\keV}, the summed spectrum from four irradiation cycles of the natural \ce{Ge} sample (timing window \qtyrange{5}{30}{\minute}) was analyzed (see \cref{fig:figure2}, bottom left). In this case, the contribution from \ce{^76Ga} is significantly lower due to its reduced isotopic abundance and because most of the \ce{^76Ga} decayed within the first \qty{5}{\minute}. The region of interest was defined as ranging from \qtyrange{2031}{2045.8}{\keV}.

To determine the average background level in this region, two constant functions were fitted to the left and right sections of the region of interest, accounting for their respective distances from the center. Energy intervals of \qtyrange{1983}{1989}{\keV} and \qtyrange{2052}{2058}{\keV} were chosen for the background fits, as these regions do not contain any known $\gamma$-ray features, such as full-energy peaks or single/double escape peaks (see \cref{tab:table1}). Only the Compton edges at \qtylist{1985.0;1985.3}{\keV}, originating from \ce{^72Ga} and \ce{^76Ga} $\gamma$-rays, respectively, may contribute to an increased background on the left side, ultimately providing a more conservative estimate for the background in the region of interest.
\begin{table}
    \caption{\label{tab:table1} Possible structures with energies $E$ in the range \qtyrange{1980}{2070}{\keV}, resulting from the neutron activation of the \ce{Ge} samples. Structures were selected if their full-energy $\gamma$-ray peaks have emission probabilities \mbox{$\nu>0.01$\%} or if their Compton edges, single, or double escape peaks exhibit \mbox{$\nu>0.1$\%}. The table also provides the corresponding emission probabilities and half-lives \cite{Singh2011NDS,Singh2010NDS,Singh2019NDS,Singh2006NDS,Singh2024NDS}.}
    \begin{ruledtabular}
        \begin{tabular}{c l l c r r}
            Channel & $E$ [keV] & $E_\gamma$ [keV] & $\nu_\gamma$ [\%] & $T_{1/2}$\\
            \hline
            \ce{^76Ge}($n,\alpha$)\ce{^73Zn} & $\,\,\,\,\,\,\,\,\,-$& 1979.7 & 0.41 & 24.5\,s \\
            \ce{^76Ge}($n,p$)\ce{^76Ga}&$\,\,\,\,\,\,\,\,\,-$ & 1980.4 & 0.19 & 30.5\,s \\
            \ce{^72Ge}($n,p$)\ce{^72Ga}& $\,\,\,\,\,\,\,\,\,-$& 1991.16 & 0.101 & 14.10\,h \\
            \ce{^74Ge}($n,p$)\ce{^74Ga}& $\,\,\,\,\,\,\,\,\,-$& 1999.3 & 0.4 & 8.12\,min \\
            \ce{^74Ge}($n,p$)\ce{^74Ga}&$\,\,\,\,\,\,\,\,\,-$ & 2004.6 & $<$0.4957 & 8.12\,min \\
            \ce{^74Ge}($n,p$)\ce{^74Ga}& $\,\,\,\,\,\,\,\,\,-$& 2014.45 & 1.29 & 8.12\,min \\
            \ce{^74Ge}($n,p$)\ce{^74Ga}&$\,\,\,\,\,\,\,\,\,-$ & 2023.9 & 0.45 & 8.12\,min \\
            \ce{^76Ge}($n,\alpha$)\ce{^73Zn}&$\,\,\,\,\,\,\,\,\,-$ & 2028.3 & 0.026 & 24.5\,s \\
            \ce{^72Ge}($n,p$)\ce{^72Ga}& $\,\,\,\,\,\,\,\,\,-$& 2028.94 & 0.115 & 14.10\,h \\
            \ce{^74Ge}($n,\alpha$)\ce{^71Zn}& $\,\,\,\,\,\,\,\,\,-$& 2064.6 & 0.045 & 2.45\,min \\
            \hline 
            \multicolumn{2}{l}{Compton edges} & & & \\ 
            \hline 
            \ce{^72Ge}($n,p$)\ce{^72Ga} & 1985.0 & 2214.024 & 0.218 & 14.10\,h \\
            \ce{^76Ge}($n,p$)\ce{^76Ga} & 1985.3 & 2214.36 & 1.98 & 30.5\,s \\
            \ce{^74Ge}($n,p$)\ce{^74Ga} & 2002.6 & 2231.9 & 0.10 & 8.12\,min \\
            \ce{^74Ge}($n,p$)\ce{^74Ga} & 2027.5 & 2257.06 & 1.75 & 8.12\,min \\
            \ce{^76Ge}($n,p$)\ce{^76Ga} & 2049.1 & 2278.80&0.39 & 30.5\,s \\
            \ce{^74Ge}($n,p$)\ce{^74Ga} & 2049.3 & 2279.05&2.34 & 8.12\,min \\
            \hline 
            \multicolumn{4}{l}{Single escape peaks $E$ = $E_\gamma$ - 511\,keV} & \\
            \hline 
            \ce{^72Ge}($n,p$)\ce{^72Ga} & 1980.026 & 2491.026& 7.73 & 14.10\,h \\
            \ce{^74Ge}($n,p$)\ce{^74Ga} & 1993.2 & 2504.2& 0.65 & 8.12\,min \\
            \ce{^72Ge}($n,p$)\ce{^72Ga} & 1996.718 &2507.718& 13.33 & 14.10\,h \\
            \ce{^72Ge}($n,p$)\ce{^72Ga} & 2004.857 &2515.857& 0.258 & 14.10\,h \\
            \ce{^76Ge}($n,p$)\ce{^76Ga} & 2013.0 & 2524.0 &0.71 & 30.5\,s \\
            \ce{^74Ge}($n,p$)\ce{^74Ga} & 2069.07 & 2580.07&1.28 & 8.12\,min \\
            \ce{^76Ge}($n,p$)\ce{^76Ga} & 2067.55 &2578.55 & 1.99 & 30.5\,s \\
            \hline 
            \multicolumn{4}{l}{Double escape peaks $E$ = $E_\gamma$ - 1022\,keV} & \\ 
            \hline 
            \ce{^74Ge}($n,p$)\ce{^74Ga} & 2008.3 & 3030.3&$<$0.1652 & 8.12\,min \\
            \ce{^74Ge}($n,p$)\ce{^74Ga} & 2009.7 & 3031.7&0.19 & 8.12\,min \\
            \ce{^76Ge}($n,p$)\ce{^76Ga} & 2012.6 & 3034.6&0.46 & 30.5\,s \\
            \ce{^76Ge}($n,p$)\ce{^76Ga} & 2047.90 & 3069.90& 0.82 & 30.5\,s \\
        \end{tabular}
    \end{ruledtabular}
\end{table}
Within the region of interest of the natural \ce{Ge} sample spectrum, a single full-energy peak is visible at \qty{2035.5+-0.4}{\keV}, attributable to \ce{^74Ga}, in agreement with the \mbox{literature \cite{Camp1971NPA1,Taylor1975CJP}}. The fit parameter for the peak width was constrained using the nearby $\gamma$-ray peak at \qty{2014.45}{\keV}. The peak is also present in the spectrum of the isotopically enriched \ce{Ge} sample, albeit with reduced statistical significance due to its lower isotopic abundance, and is absent from the spectrum of the transportation shuttle.\\

The analysis procedure was repeated using the summed spectrum of \num{50} irradiation cycles of the isotopically enriched \ce{Ge} sample, recorded by HPGe detector D1 within the \qtyrange{0}{90}{\s} timing window (see \cref{fig:figure2}, bottom right), to investigate \ce{^76Ga}-induced $\gamma$-ray peaks in the region of interest. For a direct comparison, the same energy intervals were used to approximate the background and define the region of interest as previously described for the \qtyrange{5}{30}{\minute} timing window. In this analysis, the position of the \ce{^74Ga} peak was constrained to \qty{2035.5+-0.4}{\keV} within its uncertainty to account for its significant overlap with adjacent peaks, as indicated by the dashed line in \cref{fig:figure2} (bottom right). In addition, a single fit parameter for the peak widths was applied to all peaks within the region of interest, as their energy differences are less than \qty{10}{\keV}, allowing the deviation in their energy resolution to be neglected.

In addition to the \qty{2035.5}{\keV} $\gamma$-ray peak from \ce{^74Ga}, the analysis reveals two further full-energy peaks within the region of interest at \qty{2033.1+-0.5}{\keV} and \mbox{\qty{2040.22+-0.26}{\keV}}. These features are also apparent in the accumulated spectrum recorded by HPGe detector D2 (gray) over \num{40} individual runs, confirming the presence of the three peaks at the same positions.\\

The results obtained with this Frequentist approach were compared with two alternative fitting methods.
To assess the stability of the resulting peak positions under varying background assumptions, the fit windows were extended to include regions immediately adjacent to the region of interest (\qtyrange{2027}{2050.8}{\keV}) rather than using separate background fitting regions. Notably, the background in the \qtyrange{2025}{2031}{\keV} interval for the isotopically enriched \ce{Ge} sample spectrum appears overestimated, possibly due to an underlying peak, potentially from the \ce{^73Zn} $\gamma$-ray transition. With this modified fitting procedure, the $\gamma$-ray peak of \ce{^74Ga} in the spectrum of the natural \ce{Ge} sample remains at \qty{2035.5+-0.5}{\keV} (including energy calibration uncertainties). However, the uncertainty increased by \qty{+-0.1}{\keV} owing to lower peak statistics and elevated background. Despite the higher background, the $\gamma$-ray energies and uncertainties for the \qty{2033}{\keV} and \qty{2040}{\keV} peaks remain consistent with those derived using the original method.

To independently validate the results, a Bayesian approach was employed using the Bayesian Analysis Toolkit (BAT) \cite{Caldwell2009Science}, with the fit region set to \qtyrange{2027}{2050.8}{\keV}. For the \ce{^74Ga} peak in the spectrum of the natural \ce{Ge} sample, two Gaussian priors were used: one for the peak position, based on the literature value of \mbox{\qty{2036.2+-0.4}{\keV} \cite{Singh2006NDS}}, and one for the peak width, derived from the \qty{2014.45}{\keV} peak. The Bayesian analysis reveals a peak position of \mbox{\qty{2035.8+-0.4}{\keV}} (including energy calibration uncertainties), which was then incorporated as a Gaussian prior in the triple-Gaussian fit. The final resulting peak positions from the triple-Gaussian fit are \qty{2033.1+-0.5}{\keV} and \qty{2040.19+-0.26}{\keV}.

All peak positions determined from the three analyses are consistent within their respective uncertainties, reinforcing the reliability of the results. Accordingly, the results from the initial analysis are adopted.

\section{Results}
Since the \qty{2033.1}{\keV} and \qty{2040.22}{\keV} peaks appear exclusively in the spectrum of the isotopically enriched \ce{Ge} sample within the short timing window, they are attributed to a short-lived activation product of \ce{^76Ge}. Based on reaction cross sections, half-lives, and $Q$ values, the only viable candidates are \ce{^73Zn} and \ce{^76Ga}. However, a previous experiment by Vedia \textit{et al.} \cite{Vedia2017PRC}, which investigated the decay spectrum of \ce{^73Zn}, reported no evidence of a corresponding $\gamma$-ray peak in the region of interest. This observation supports the conclusion that both the \qty{2033.1}{\keV} and \qty{2040.22}{\keV} peaks originate from \ce{^76Ga}.

Within a $2\sigma$ peak interval, a total of \num{278} counts were recorded for the \qty{2035.5}{\keV} peak in the natural \ce{Ge} spectrum (timing window \qtyrange{5}{30}{\minute}), which exceeds the critical limit of \num{109} counts established at a one-tailed confidence level of 99.87\%. For the spectrum of the isotopically enriched \ce{Ge} sample (timing window \qtyrange{0}{90}{\s}), where significant overlap between peaks is present, the net count areas were determined by calculating the fractional contribution of each peak’s amplitude to the total net counts within the $2\sigma$ interval (starting from $-2\sigma$ relative to the \qty{2033.1}{\keV} peak center). In this case, \num{329} counts were attributed to the \qty{2033.1}{\keV} peak, surpassing its critical limit of \num{213} counts (99.87\% C.L.), and \num{744} counts were measured for the \qty{2040.22}{\keV} peak, far exceeding its critical limit of \num{200} counts (99.87\% C.L.). For a conservative estimate, the counts from the two overlapping peaks were also included as background when calculating the critical limit for a single $\gamma$-ray peak.
\begin{figure*}
		\includegraphics[width=\textwidth, trim=32mm 190mm 22mm 25mm]{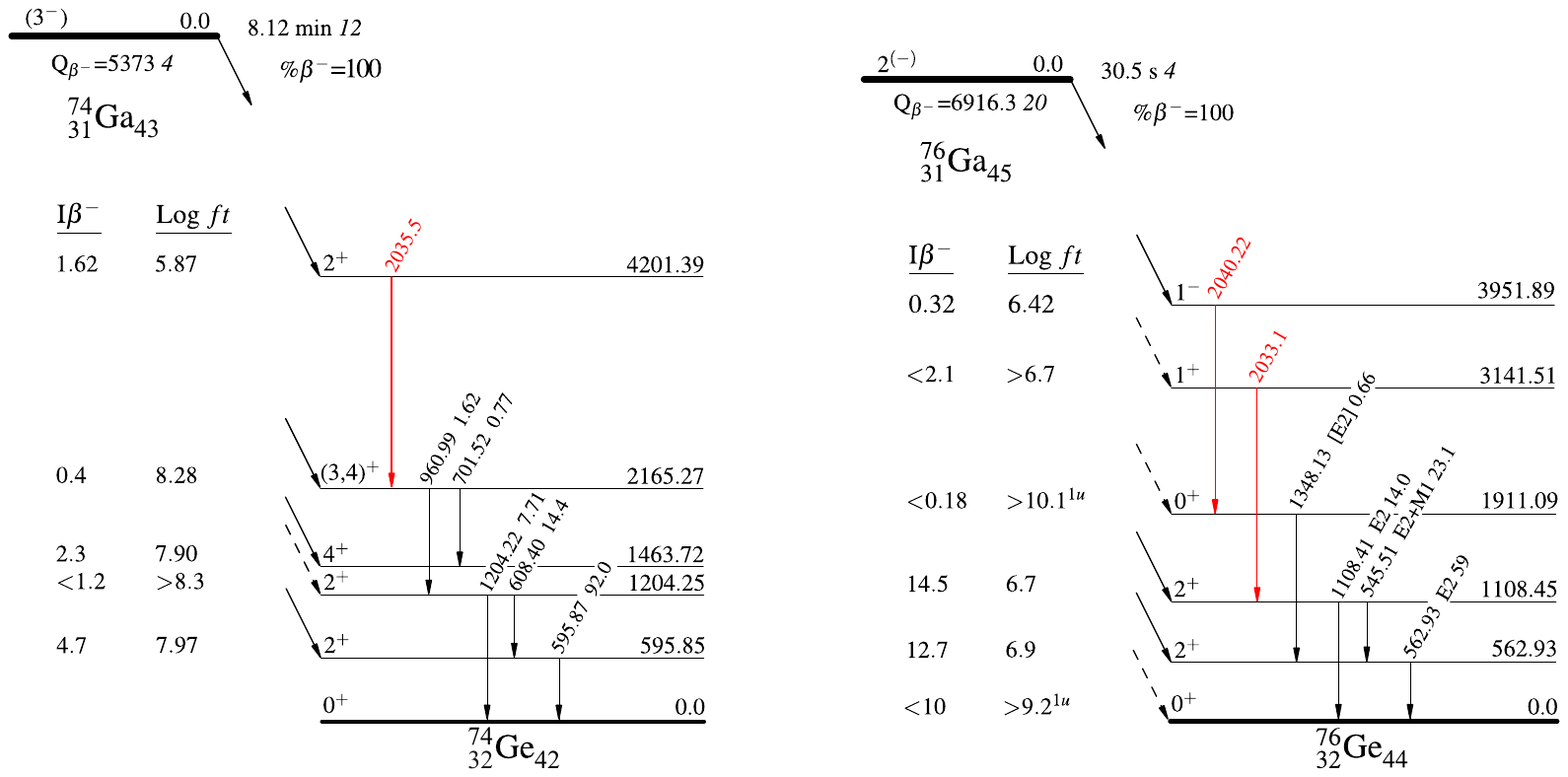}%
	\caption{\label{fig:figure3} 
    Decay schemes of \ce{^74Ga} and \ce{^76Ga} illustrating the presumed origins of the three $\gamma$ rays detected near \qty{2039}{\keV} (highlighted in red). Data are based on the decay information reported in \cite{Singh2006NDS,Singh2024NDS}.}
\end{figure*} 
\section{Conclusions}
More than fifty years have passed since the first experimental indications of $\gamma$ rays from neutron activation of germanium within the signal region for \ce{^76Ge} $0\nu\beta\beta$ decay were reported. Here, these earlier findings were re-evaluated, and three $\gamma$-ray peaks near \qty{2039}{\keV} have been detected following fast neutron activation of germanium isotopically enriched in \ce{^76Ge}: a $\gamma$-ray peak at \qty{2035.5+-0.4}{\keV} attributed to \ce{^74Ga} and two $\gamma$-ray peaks at \qty{2033.1+-0.5}{\keV} and \qty{2040.22+-0.26}{\keV}, both associated with \ce{^76Ga}. This observation indicates that germanium itself may contribute to the background in \ce{^76Ge} $0\nu\beta\beta$ decay experiments such as LEGEND. However, since all three $\gamma$-ray transitions are likely to occur as cascades, the high-efficiency veto in the LEGEND experiment is expected to mitigate this background. The corresponding decay scheme giving rise to these three $\gamma$-ray transitions is illustrated in \cref{fig:figure3}.

The $\gamma$ ray at \qty{2035.5}{\keV} originating from \ce{^74Ga} was remeasured in the current experiment and found to be in agreement with previous reports \cite{Camp1971NPA1,Taylor1975CJP}. No improvement in precision was achieved.

In addition, a newly detected \qty{2033.1}{\keV} $\gamma$-ray emitted during \ce{^76Ga} decay may be induced by the de-excitation of the previously identified \qty{3141.51+-0.07}{\keV} level (fed by less than 2.1\% via $\beta$ decay) to the lower-lying \qty{1108.45+-0.04}{\keV} level. This possibility will be further explored in an additional coincidence measurement with the isotopically enriched \ce{Ge} sample placed between two HPGe detectors.

Moreover, the transition between the energy levels at \qtylist{3951.89;1911.09}{\keV} in \ce{^76Ge} is considered to be the origin of the \qty{2040.22+-0.26}{\keV} $\gamma$-ray peak. This result, originally measured by Camp and Foster \cite{Camp1971NPA2}, is confirmed here within a $2\sigma$ agreement.

Finally, the observations reported by the Kentucky group \cite{Crider2015PRC,Mukhopadhyay2017PRC}, which claimed a \qty{2038}{\keV} $\gamma$-ray peak arising from the de-excitation of a newly discovered \qty{3147.28}{\keV} level, could not be reproduced. This discrepancy may stem from insufficient feeding of that level by $\beta$ decay, rendering the associated \qty{2038.89}{\keV} transition unobservable in the current experiment.

\begin{acknowledgments}
The authors thank B. Lehnert for his insightful feedback and A. Hartmann for his valuable technical support. Financial support provided by the Bundesministerium für Bildung und Forschung (BMBF-Projekt 05A17OD1) is gratefully acknowledged.
\end{acknowledgments}
%
\def\AAA{Astro. and Astrophys.}
\def\AAS{Astro. and Astrophys. Suppl.}
\def\AHEP{Adv. High Energy Phys.}
\def\AIPCP{AIP Conf. Proc.}
\def\ANDT{Atom. Nucl. Data Tab.}
\def\ANDT{Atomic and Nuclear Data Tables}
\def\APB{{Acta Pol.} B}
\def\APJ{ApJ}
\def\arxiv{arXiv}
\def\APJ{The Astrophys. J.}
\def\APJL{The Astrophys. J. Lett.}
\def\APP{Astropart. Phys.} 
\def\ARI{Appl. Rad. Isot.}
\def\ARNPS{Ann. Rev. Nucl. Part. Sci.}
\def\CJP{Can. J. Phys.}
\def\CPC{Chin. Phys. C}
\def\CPCo{Comput. Phys. Commun.}
\def\EPJA{Europ. Phys. J. A}
\def\EPJC{Europ. Phys. J. C}
\def\EPJP{Europ. Phys. J. Plus}
\def\EPJWC{EPJ Web conf.}
\def\EPL{Europhys. Lett.}
\def\ITNS{IEEE Trans. Nucl. Sci.}
\def\JCAP{JCAP}
\def\JINST{JINST}
\def\JPCS{J. Phys. Conf. Ser.}
\def\JPG{J. Phys. G}
\def\MNRAS{Month Not. Royal Ast. Soc.}
\def\MUP{Moscow Univ. Phys.}
\def\NAF{Z. Naturforschung A} 
\def\NAT{Nature}
\def\NCI{Nuovo Cimento C}
\def\NDS{Nucl. Data Sheets}
\def\NIM{Nucl. Instrum. Methods}
\def\NIMA{{Nucl. Instrum. Methods} A}
\def\NP{Nucl. Phys.}
\def\NPA{{Nucl. Phys.} A} 
\def\NPB{{Nucl. Phys.} B} 
\def\NPBP{{Nucl. Phys.} B (Proc. Suppl.)} 
\def\PAC{Pure Appl. Chem.}
\def\PAN{{Physics of Atomic Nuclei}} 
\def\PHM{Philos. Mag.} 
\def\PLB{{Phys. Lett.} B}
\def\PNpp{Prog. Nucl. Part. Phys.}  
\def\PR{Phys. Rev.} 
\def\PRC{{Phys. Rev.} C} 
\def\PRD{{Phys. Rev.} D} 
\def\PRL{Phys. Rev. Lett.} 
\def\PRP{{Phys. Rep.}}
\def\PTEP{Prog. Th. Exp. Phys.}
\def\PTP{{Prog. Theo. Phys.}}
\def\RMP{{Rev. Mod. Phys.}}
\def\Rpp{Rep. Prog. Phys.}
\def\SCI{Science} 
\def\ZP{Z. Phys.} 
\def\ZPA{{Z. Phys.} A} 
\def\ZPC{{Z. Phys.} C} 

\end{document}